\documentclass[pra,aps,a4paper,twocolumn,repreprint,superscriptaddress,longbibliography]{revtex4}

\usepackage{physics}
\usepackage{amssymb}
\usepackage{amsmath}
\usepackage{amsfonts}
\usepackage{hyperref}
\usepackage{graphicx}
\usepackage[dvipsnames]{xcolor}
\usepackage{bm}

\usepackage{qcircuit}

\definecolor{mygold}{rgb}{0.93,0.69,0.13}

\definecolor{mypurple}{rgb}{0.49,0.18,0.56}

\newcommand{\hc}{\mathop{\rm h.c.}}

\def\be{\begin{equation}}
\def\ee{\end{equation}}

\begin{document}

\title{Quantum simulation of the 1D Fermi-Hubbard model as a $\mathrm{Z}_2$ lattice-gauge theory}

\author{Uliana E. Khodaeva}
\affiliation{Technical University of Munich, TUM School of Natural Sciences, Physics Department, 85748 Garching, Germany}

\author{Dmitry L. Kovrizhin}
\affiliation{LPTM, CY Cergy Paris Universite, UMR CNRS 8089, Pontoise 95032 Cergy-Pontoise Cedex, France}

\author{Johannes Knolle}
\affiliation{Technical University of Munich, TUM School of Natural Sciences, Physics Department, 85748 Garching, Germany}
\affiliation{Munich Center for Quantum Science and Technology (MCQST), Schellingstr. 4, 80799 M{\"u}nchen, Germany}
\affiliation{Blackett Laboratory, Imperial College London, London SW7 2AZ, United Kingdom}

\begin{abstract}
The Fermi-Hubbard model is one of the central paradigms in the physics of strongly-correlated quantum many-body systems. Here we propose a
quantum circuit algorithm based on the $\mathrm{Z}_2$ lattice gauge theory (LGT) representation of the one-dimensional Fermi-Hubbard model, which is suitable
for implementation on current NISQ quantum computers. Within the LGT description there is an extensive number of local conserved quantities
commuting with the Hamiltonian. We show how these conservation laws can be used to implement an efficient error-mitigation scheme. The latter is based
on a post-selection of states for noisy quantum simulators. While the LGT description requires a deeper quantum-circuit compared to a Jordan-Wigner (JW)
based approach, remarkably, we find that our error-correction protocol leads to results being on-par with a standard JW implementation
on \textit{noisy} quantum simulators.

\end{abstract}

\date{\today} 

\maketitle

\section{Introduction}

In recent years there has been remarkable progress in the development of new quantum computation technologies. The availability of 
NISQ quantum computers in combination with new quantum computation algorithms has offered new possibilities for
simulations of quantum many-body systems~\cite{preskill2018quantum,boixo2018characterizing}. While the technology is arguably in its infancy, there is a growing effort in the development of new computational algorithms which could help to mitigate the inevitable presence of noise and errors on currently available quantum hardware~\cite{Smith2019,kandala2019error,Li_2017,He_2020}. Recent research in this field resulted in new efficient error-mitigation schemes~\cite{temme2017error,bravyi2021mitigating,vovrosh2021simple,nation2021scalable,huggins2021virtual, Sopena_2021, Funcke_2022}, bringing NISQ forward as a powerful 
tool for studying many-body quantum problems. Most prominent is the simulation of non-equilibrium problems, which are
otherwise very hard to study using other theoretical and numerical techniques. Some remarkable examples
of recent achievements include quantum simulations of time-crystals~\cite{Mi2022}, lattice gauge theories~\cite{martinez2016real,yang2020observation,PRXQuantum.3.020320}, confinement dynamics~\cite{tan2021domain,vovrosh2021confinement,Mildenberger2022} and quantum spin-liquids showing topological order~\cite{satzinger2021realizing}. 

One of the simplest, yet one of the most versatile, examples of a strongly-interacting many-body
quantum systems is given by the Fermi-Hubbard model (FHM)~\cite{hubbard1964electron}. It has been used as a minimal model for a large variety of physical systems, from high-T$_c$ superconductors to cold atomic gases in optical lattices~\cite{arovas2022hubbard}. However, exact results for this model are restricted to isolated fine-tuned points of the parameter space~\cite{lieb1989two,tasaki1998hubbard}, and except for the one-dimensional case 1D~\cite{Essler, giamarchi2003quantum} understanding its physics remains one of the grand challenges of condensed matter physics. In this context, the Fermi-Hubbard model has been extensively studied and benchmarked with numerical methods~\cite{scalapino2007numerical,leblanc2015solutions,qin2022hubbard}, and recently algorithms for digital quantum simulations have been discussed~\cite{wecker2015progress,jiang2018quantum,cade2020strategies,cai2020resource,stanisic2022observing}. 

In this paper we propose and benchmark a new quantum simulation protocol for the Fermi-Hubbard model which can be implemented
on NISQ quantum computers. Our approach is based on a mapping to a $\textrm{Z}_2$ lattice-gauge theory, which describes a system
of lattice fermions coupled to spin-1/2 degrees of freedom defined on the links of the lattice~\cite{ruegg2010z,fabrizio2015z}. This mapping has been used in
quantum Monte-Carlo studies of the 2D Hubbard model with attractive interactions, where it allows one to use determinant QMC without the sign problem~\cite{hohenadler2018fractionalized,hohenadler2019orthogonal}, and played an important role in the discovery of the physics of disorder-free localisation~\cite{smith2017disorder,Smith2018,papaefstathiou2020disorder}. In the context of quantum simulation the implementation of LGTs itself has seen many recent advances~\cite{Bender_2018,TAGLIACOZZO2013160,banuls2020review,bauer2023quantum, Huffman_2022}.

The LGT description of the Fermi-Hubbard model is obtained by introducing auxiliary spin-1/2 degrees of freedom, so
one has to introduce constraints on every lattice site in order to remove the redundancy of the description. These constraints
fix the eigenvalues of a set of local operators to $\pm1$ depending on the sign of the interaction term in the Hubbard model.
These operators are further conserved under the time evolution with the LGT Hamiltonian. From a quantum simulation
perspective, the LGT representation allows us to propose and implement an efficient error-correction protocol based on
post-selection of states utilising a macroscopic number of these conserved quantities.

We focus on studying the real time quench dynamics using the LGT mapping for the 1D Hubbard model,
and compare these results with the ones obtained by a direct simulation of the FHM using a spin-1/2
representation (via Jordan-Wigner transformations). Remarkably, in the presence of the noise we find that
despite the fact that the LGT approach requires a much larger number of CNOT gates compared with the
direct simulation (6 gates vs 2 gates per hopping term) it offers results whose error is on par or
better than that obtained from a direct simulation. We note that as long as the number of simulation runs
is not an issue, the power of our error-mitigation scheme based on local conserved quantities
of the LGT can outperform noisy simulations utilising only global conservation laws. This unexpected observation suggests
further exploration of the LGT approach to models of strongly-correlated systems, e.g. a 2D FHM.

The structure of the paper is the following. In section II we define the model and its mapping to a $\mathrm{Z}_2$ LGT.
In section III we show how to implement the LGT in terms of quantum circuits with a minimum number of CNOT gates, and explain
in particular the quantum-circuit representation of the three-spin coupling terms. In section IV we describe the details of
our error-mitigation protocol. The results of the simulations are discussed in section V.

\section{The 1D Fermi-Hubbard model and its mapping to a $\mathrm{Z}_2$ LGT}
We consider a system of spin-1/2 lattice fermions on a general lattice in any dimensions. The fermion operators are defined by $\hat{c}_{i\sigma}$, where $i,\sigma$
are the site and spin indices correspondingly, and obey standard anti-commutation relations $\{\hat{c}_{i\sigma},\hat{c}^{\dagger}_{j\sigma'}\}=\delta_{i j}\delta_{\sigma \sigma'}$. The kinetic energy is given by the nearest-neighbour hopping term with strength $J$,
and the Hubbard interaction strength is denoted by $U$,
\be\label{hamilt_hub}
\hat{H} = -J\sum_{\langle ij\rangle\sigma}(\hat{c}_{i\sigma}^{\dagger}\hat{c}_{j\sigma} + h.c.) + \frac{U}{2}\sum_{i}(\hat{n}_i - 1)^2.
\ee
Here the summation in the first term on the r.h.s goes over nearest-neighbour sites, the operator \mbox{$\hat{n}_i = \sum_{\sigma}\hat{c}_{i\sigma}^{\dagger}\hat{c}_{i\sigma}$} describes the total number of fermions on site $i$.

Let us first present the details of the mapping of the Hamiltonian (\ref{hamilt_hub}) to a $\mathrm{Z}_2$ LGT, where we follow the slave-spin representation of Ref.~\cite{ruegg2010z}. We introduce pseudospin operators $\hat{I}$ on each site, having eigenstates $\hat{I^z}|\pm\rangle = \pm\frac{1}{2}|\pm\rangle$, where the state $|+\rangle$ with a positive eigenvalue denotes both doubly occupied and empty states on the lattice, and $|-\rangle$ corresponds to states with a single fermion. In other words, the eigenvalues of these operator are in one to one correspondence with the absence/presence of a local magnetic moment. One has to also define another set of spin-1/2 fermion operators $\hat{f}_{i\sigma}$ on each lattice site. In terms of these operators and the pseudospins the physical fermion operators $\hat{c}_{i\sigma}^{\dagger}$ can then be written as $\hat{c}_{i\sigma}^{\dagger} = 2\hat{I}^{x}_{i}\hat{f}_{i\sigma}^{\dagger}$.
The FHM Hamiltonian (\ref{hamilt_hub}) in slave-spin representation now reads,
\be\label{slave_spin}
\hat{\tilde{H}} = -4J\sum_{\langle ij\rangle\sigma}(\hat{I}_i^{x}\hat{I}_j^{x}\hat{f}_{i\sigma}^{\dagger}\hat{f}_{j\sigma} + \hc) +  \frac{U}{2}\sum_{i}(\hat{I}_i^{z} + 1/2).
\ee
Because the slave-spin representation requires an enlarged Hilbert space, one has to introduce constraints. Therefore, the Hamiltonian \ref{slave_spin} is equivalent to the FHM Hamiltonian \eqref{hamilt_hub} only in the physical subspace, i.e. when the operators satisfy the following equations
\be\label{phys_constraint}
\hat{I}_i^z + \frac{1}{2} - (\hat{n}_i - 1)^2 = 0,~~\forall i.
\ee
Using this representation one can now transform the Hamiltonian into a LGT form, where the next step is to carry out a duality transformation of the operators $\hat{I}^z_i$ by  introducing spin-1/2 operators $\hat{\tau}_{jk}$ living on the bonds between adjacent sites,
\be
\hat{\tau}^{z}_{jk} = \hat{I}^{x}_{j}\hat{I}^{x}_{k},~~\hat{A}_{j} = \hat{I}_{j}^{z}.
\ee
It is worth noting that all the steps above are valid in arbitrary dimension, and we can write $\hat{A}_{j} = \prod_{i \in \textit{star(j)}}\hat{\tau}^x_{ij}$. Here the ``star'' of the lattice site $j$ denotes the set of the sites being nearest neighbours of the site $j$ connected to it by a hopping term. In the 1D case this simply becomes
\be
\hat{I}_{j}^{z} = \hat{\tau}^x_{j-1,j}\hat{\tau}^x_{j,j+1}.
\ee
After performing this duality transformation, we arrive at the following $\mathrm{Z}_2$ LGT Hamiltonian which we write below for the 1D case,
\begin{multline}\label{hamilt_LGT}
\hat{H}_{LGT} = -4J\sum_{j, \sigma}(\hat{\tau}^{z}_{j,j+1}\hat{f}_{j\sigma}^{\dagger}\hat{f}_{j+1 \sigma} + \hc) \\+ \frac{U}{2}\sum_j \hat{\tau}_{j-1,j}^x\hat{\tau}_{j,j+1}^x,
\end{multline}
where we omitted a constant $UN/2$ term because it does not affect the time evolution of the system. 

\begin{figure}[h!]\label{schematics}
\includegraphics[width=8.0cm]{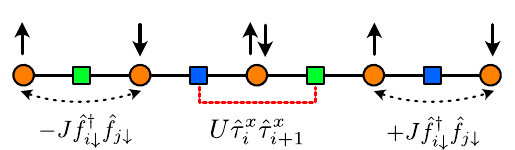}
\caption{Schematic picture of the mapping to a LGT \eqref{hamilt_LGT}. In the FHM model the spin-1/2 fermions live on the sites of the lattice (circles) connected by hopping, and the Hubbard term contributes to the energy whenever there are two fermions with opposite spins on the same lattice site. In the lattice gauge theory description we have additional spin-1/2 degrees of freedom on every link (squares). The sign of the fermion hopping is given by the direction of bond spins (green - spin up, blue - spin down) in the z-representation. The  bond spins interact via transverse Ising interactions (red dotted line).} \label{schematic}
\end{figure}

The Hamiltonian now describes spin-1/2 fermions hopping on a lattice with the sign of the hopping defined by the eigenvalue of the $\tau^{z}_{ij}$, see Fig.1. The bond spins interact via a transverse-field Ising term. In two dimensions the LGT Hamiltonian is related to the Toric-code coupled to fermions with hopping, under additional constraints arising from the projection to the physical subspace. The defining feature of the Hamiltonian \eqref{hamilt_LGT} is the presence of a local $\textrm{Z}_2$ gauge invariance, namely that the fermion operators are invariant under the following transformations $\hat{f}_{i\sigma} \rightarrow \theta_i \hat{f}_{i\sigma}$ and $\hat{\tau}_{i,i+1}^z \rightarrow \theta_i \hat{\tau}_{i,i+1}^z \theta_{i+1}$ with $\theta_i = \pm 1$. Therefore, this model is indeed an example of a $\textrm{Z}_2$ LGT.

It is important to note, that because we employed a duality transformation, we have to select one of the disconnected subspaces of the model, where every such subspace is enumerated by a set of conserved quantities defined by the products of $\hat{\tau}^z$ along closed loops. Since here we study a 1D model with periodic boundary conditions, the only operator of this type is the following Wilson loop
\be\label{gamma}
\hat{\Gamma} = \prod_{i=1}^{N}\hat{\tau}^z_{i,i+1},
\ee
where $N$ is the number of sites. This operators has eigenvalues $\pm 1$, and since the choice of a corresponding sector is arbitrary, we fix it to be $\hat{\Gamma} = +1$ in the calculations.

\section{Quantum circuit representation}
\subsection{Fermion encoding}
In order to encode both the fermionic Hamiltonian \eqref{hamilt_hub} and the LGT Hamiltonian \eqref{hamilt_LGT} in terms of qubits, we first have to represent fermion operators as spin-1/2. In the 1D case this can be done using the Jordan-Wigner (JW) transformation. In order to preserve the canonical anti-commutation relations between the operators related to fermions with different spins, we are going to employ the following version of the JW-transformation~\cite{shastry1986infinite},
\be\label{JW}
\begin{split}
&\hat{f}^{\dagger}_{i\uparrow} = \hat{K}_{i\uparrow}\hat{S}^{+}_{i\uparrow},~~\hat{f}_{i\uparrow} = \hat{K}_{i\uparrow}\hat{S}^{-}_{i\uparrow},\\
&\hat{f}^{\dagger}_{i\downarrow} = \hat{P}_{\uparrow}\hat{K}_{i\downarrow}\hat{S}^{+}_{i\downarrow},~~\hat{f}_{i\downarrow} = \hat{P}_{\uparrow}\hat{K}_{i\downarrow}\hat{S}^{-}_{i\downarrow},
\end{split}
\ee
and $\hat{n}_{i\sigma} = \hat{S}^z_{i\sigma} + 1/2$. Here $\hat{S}_{i,\uparrow}$ and $\hat{S}_{i,\downarrow}$ are two flavours of spin-1/2 operators, which encode spin-up and spin-down fermions correspondingly, and we also introduced string-operators $\hat{K}_{i\sigma} = e^{i\pi\sum_{k = 1}^{i-1}\hat{n}_{k\sigma}}$ as well as spin-up fermion parity operator $\hat{P}_{\uparrow} = e^{i\pi\sum_{k=1}^{N}\hat{n}_{k\uparrow}}$, whose eigenvalues are $+1$ and $-1$ when the total number of spin-up fermions is even/odd correspondingly. 

After application of the Jordan-Wigner transformation we obtain the spin representation of the Hubbard Hamiltonian \eqref{hamilt_hub},
\be\label{hamilt_hub_spin_1}
\hat{H} =-J\sum_{j\sigma}(\hat{S}^{+}_{j\sigma}\hat{S}^{-}_{j+1\sigma} + \hc)+ \frac{U}{2}\sum_j(\hat{S}^{z}_{j\uparrow} + \hat{S}^{z}_{j\downarrow})^2.
\ee
Taking into account the conservation of up-spin and down-spin magnetization, which follows from the corresponding particle conservation laws, one can further simplify this expression, and we obtain
\be\label{hamilt_hub_spin}
\hat{H}= -J\sum_{j,\sigma}(\hat{S}^{+}_{j\sigma}\hat{S}^{-}_{j+1\sigma} + \hc) + U\sum_j \hat{S}^{z}_{j\uparrow} \hat{S}^{z}_{j\downarrow}.
\ee

Having carried out the same JW transformation for the LGT Hamiltonian, we obtain a representation of (\ref{hamilt_LGT}) in terms of spin operators
\begin{multline}\label{hamilt_lgt_spin}
\hat{H}_{LGT} = -4J\sum_{j, \sigma}(\hat{\tau}^{z}_{j,j+1}\hat{S}_{j\sigma}^{+}\hat{S}_{j+1 \sigma}^{-} + \hc) \\+ \frac{U}{2}\sum_j \hat{\tau}_{j-1,j}^x\hat{\tau}_{j,j+1}^x.
\end{multline}

It is worth noting that the boundary conditions for the spin representation depend on the parity of the total number of fermions. Namely, for even total number of fermions periodic boundary conditions (PBC) transform into anti-periodic boundary conditions (APBC), while in the case of odd total number of fermions, APBC transform into PBC.

\subsection{Implementation of the time evolution}
Our goal is the quantum simulation of a quench dynamics, see e.g.~\cite{Smith2019}.  We implement the time evolution under the Hamiltonians discussed above using the standard Suzuki-Trotter decomposition. In other words we discretize time by splitting the entire time evolution into $N = t/\Delta t$ steps with the fixed length $\Delta t$. For the evolution operator $\hat{U} = e^{-i\hat{H}t}$, this procedure gives $\hat{U}(t) = (\hat{U}(\Delta t))^{N}$. Each application of $\hat{U}(\Delta t)$ is referred to as a Trotter step.

Now one can approximate the evolution operator for each Trotter step by the sequence of relatively simple operators. Let's consider a Hamiltonian of the form $\hat{H} = \hat{A} + \hat{B}$, where $[\hat{A}, \hat{B}] \neq 0$. The application of the Trotter decomposition is based on the following expansion for small $\Delta t$,
\be
e^{-i\hat{H}\Delta t} = e^{-i\hat{A}\Delta t}e^{-i\hat{B}\Delta t} + O[(\Delta t)^2]
\ee
but in practise one can use sizeable step sizes to reach longer times~\cite{ciavarella2022floquet}.

Both Hamiltonians $\hat{H}$ and $\hat{H}_{LGT}$ are naturally divided into two non-commuting terms, a hopping term and an on-site interaction term. Therefore, the time evolution governed by the Hubbard Hamiltonian \eqref{hamilt_hub_spin} can be approximated in the following way,
\begin{multline}
e^{-i\hat{H}\Delta t} \approx\prod_{j_{\text{even}}\uparrow}\hat{A}_{j\uparrow}\times\prod_{j_{\text{odd}}\uparrow}\hat{A}_{j\uparrow}\times \prod_{j_{\text{even}}\downarrow}\hat{A}_{j\downarrow}\times\prod_{j_{\text{odd}}\downarrow}\hat{A}_{j\downarrow}\\
\times\prod_{j}\hat{B}_j + O[(\Delta t)^2],
\end{multline}
where we have introduced the following operators
\be
\hat{A}_{j\sigma} = e^{iJ\Delta t(\hat{S}^{+}_{j\sigma}\hat{S}^{-}_{j+1\sigma} + \hc)},\ \ \hat{B}_j = e^{-iU\Delta t \hat{S}^{z}_{j\uparrow} \hat{S}^{z}_{j\downarrow}}.
\ee

In a similar way we can write the corresponding trotterized time evolution operator for the LGT Hamiltonian \eqref{hamilt_lgt_spin} for a single time step as,
\begin{multline}
e^{-i\hat{H}_{LGT}\Delta t} =\prod_{j_{\text{even}}\uparrow}\hat{C}_{j\uparrow}\times \prod_{j_{\text{odd}}\uparrow}\hat{C}_{j\uparrow}
\times\prod_{j_{\text{even}}\downarrow}\hat{C}_{j\downarrow}\times\prod_{j_{\text{odd}}\downarrow}\hat{C}_{j\downarrow}\\
\times\prod_{j_{\text{even}}}\hat{D}_j\times\prod_{j_{\text{odd}}}\hat{D}_j+ O[(\Delta t)^2],
\end{multline}
where we have introduced the following operators:
\be
\begin{split}
&\hat{C}_{j\sigma} = e^{4iJ\Delta t(\hat{\tau}^z_{j,j+1}\hat{S}^{+}_{j\sigma}\hat{S}^{-}_{j+1\sigma} + \hc)}\\
&\hat{D}_j = e^{-i(U/2)\Delta t \hat{\tau}^{x}_{j-1,j} \hat{\tau}^{x}_{j,j+1}}.
\end{split}
\ee
\subsection{Quantum Circuits}
The final step towards the implementation of the time evolution of the aforementioned systems on a quantum simulator is to express both the states and the operators in terms of quantum circuits. The quantum circuit representation is therefore given by an array of qubits and a series of one- and two-qubit quantum gates~\cite{nielsen2002quantum}. The computational basis is taken as a tensor product of single-qubit states in the $z$-representation.

In this work, we provide an implementation based on the IBM quantum computer architecture. Therefore, we implement the quantum circuit using CNOT gates, which reverse the state of the second qubit depending on the state of the first one, as the only gates which generate entanglement,
\be
\textrm{CNOT}=
\Qcircuit @C=.5em @R = .5em{
& \ctrl{1} & \qw\\
& \targ & \qw } =
\begin{pmatrix}
1 & 0 & 0 & 0\\
0 & 1 & 0 & 0\\
0 & 0 & 0 & 1\\
0 & 0 & 1 & 0\\
\end{pmatrix}.
\ee
We also employ the following single-qubit gates, which are provided by the IBM quantum computing architecture:

-- Pauli matrices:
\be
\begin{split}
\Qcircuit @C=.5em @R = .5em{
& \gate{X} & \qw
} &= 
\begin{pmatrix}
0 & 1\\
1 & 0\\
\end{pmatrix},~~\Qcircuit @C=.5em @R = .5em{
 & \gate{Y} & \qw
} = 
\begin{pmatrix}
0 & -i\\
i & 0\\
\end{pmatrix}\\
\Qcircuit @C=.5em @R = .5em{
 & \gate{Z} & \qw
} &= 
\begin{pmatrix}
1 & 0\\
0 & -1\\
\end{pmatrix},
\end{split}
\ee

-- Phase gates:
\be
\Qcircuit @C=.5em @R = .5em{
& \gate{H} & \qw
} = \frac{1}{\sqrt{2}}
\begin{pmatrix}
1 & 1\\
1 & -1\\
\end{pmatrix},~~
\Qcircuit @C=.5em @R = .5em{
 & \gate{S} & \qw
} =
\begin{pmatrix}
1 & 0\\
0 & i\\
\end{pmatrix},
\ee
(the Hadamard and the S-phase gate)

-- Rotation gates:
\begin{equation}
\begin{split}
\Qcircuit @C=.5em @R = .5em{
& \gate{R_{X}(\theta)} & \qw} 
&= 
\begin{pmatrix}
\cos\frac{\theta}{2} & -i\sin\frac{\theta}{2}\\
-i\sin\frac{\theta}{2} & \cos\frac{\theta}{2}\\
\end{pmatrix}\\
\Qcircuit @C=.5em @R = .5em{
 & \gate{R_{Y}(\theta)} & \qw
} &= 
\begin{pmatrix}
 \cos\frac{\theta}{2} & -\sin\frac{\theta}{2}\\
 \sin\frac{\theta}{2} & \cos\frac{\theta}{2}\\   
\end{pmatrix}\\
\Qcircuit @C=.5em @R = .5em{
 & \gate{R_{Z}(\theta)} & \qw
} &= 
\begin{pmatrix}
 e^{-i\frac{\theta}{2}} & 0\\
 0 &  e^{i\frac{\theta}{2}}\\   
\end{pmatrix}.
\end{split}
\end{equation}

The optimal decomposition of arbitrary two-qubit gates~\cite{vatan2004optimal} can be used to express the operators $\hat{A}$, $\hat{B}$ and $\hat{D}$. The hopping operator $\hat{A}$ related to the Hubbard Hamiltonian can be written as the following circuit
\be
 \Qcircuit @C=.5em @R = 1em{
&\lstick{S_{i\sigma}} & \multigate{1}{F}  & \gate{R_y(+J\Delta t/2)} &  \multigate{1}{G} & \qw\\
&\lstick{S_{i+1\sigma}} & \ghost{F}  & \gate{R_y(-J\Delta t/2)} &  \ghost{G} & \qw
    }
\ee
where the blocks $F$ and $G$ denote the circuits
\be \small
\Qcircuit @C=.5em @R = 1em{
& \multigate{1}{F} & \qw\\
 & \ghost{F} & \qw
 }  =\hspace{0.5 cm} \Qcircuit @C=.5em @R = 1em{
&\lstick{S_{i\sigma}} & \gate{H} & \gate{S} & \gate{H} & \ctrl{1} & \qw &\qw & \gate{S} & \gate{H} & \qw\\
&\lstick{S_{i+1\sigma}} & \gate{H}  & \gate{S} & \gate{H} & \targ & \gate{H} & \gate{Z} & \gate{S} &\gate{H} & \qw
 }
\ee
\be
\small
\Qcircuit @C=.5em @R = 1em{
& \multigate{1}{G} & \qw\\
 & \ghost{G} & \qw
 }  =\hspace{0.5 cm} \Qcircuit @C=.5em @R = 1em{
&\lstick{S_{i\sigma}} & \gate{H} & \gate{Z} & \gate{S} &  \ctrl{1} &  \gate{H} & \gate{Z} & \gate{S} & \gate{H} & \qw \\
&\lstick{S_{i+1\sigma}} & \gate{H} & \gate{S} & \gate{H} & \targ  &  \gate{H} & \gate{Z} & \gate{S} & \gate{H} & \qw
 }
\ee
The operator $\hat{B}$ related to the on-site interaction term in the Hubbard Hamiltonian can be written as
\be
\Qcircuit @C=.5em @R = 1em{
\lstick{S_{j\uparrow}} & \ctrl{1} & \qw & \ctrl{1} & \qw\\
\lstick{S_{j\downarrow}} & \targ & \gate{R_z(2U\Delta t)} & \targ & \qw
}
\ee
Finally, the operator $\hat{D}$ describing the interaction of the bond spins in the LGT Hamiltonian is written as
\be
\Qcircuit @C=.5em @R = 1em{
\lstick{\tau_{j,j-1}} & \ctrl{1} & \gate{R_x(U\Delta t)} & \ctrl{1} & \qw\\
\lstick{\tau_{j,j+1}} & \targ & \qw & \targ & \qw
}
\ee
In order to implement the three-qubit operator $\hat{C}$, which is related to the fermion hopping term in the LGT Hamiltonian, we have
constructed the following minimal quantum circuit (with the blocks $F$ and $G$ defined above),
\be
 \Qcircuit @C=.5em @R = 1em{
&\lstick{\tau_{i,i+1}} & \qw & \ctrl{1} & \ctrl{2} & \qw & \ctrl{2} & \ctrl{1} & \qw & \qw\\
&\lstick{S_{i\sigma}} & \multigate{1}{F} & \targ & \qw & \gate{R_y(+J2\Delta t)} & \qw & \targ & \multigate{1}{G} & \qw\\
&\lstick{S_{i+1\sigma}} & \ghost{F} & \qw & \targ & \gate{R_y(-2J\Delta t)} & \targ & \qw & \ghost{G} & \qw
    }
\ee
and we note that this representation of the three-spin coupling term requires 6 CNOT gates in total. One can compare this with
the representation of two-spin interactions, which requires only 2 CNOT gates.

\subsection{Resource estimation}
According to their name, the Noisy Intermediate Scale Quantum (NISQ) devices are currently suffering from considerable errors and for the moment still have a relatively small number of qubits. One of the main sources of errors are the gate errors. Since the two-qubit error rate tends to be an order of magnitude above the single qubit error rate, the number of CNOTs can be used as a representative of these errors.

Above we have presented two approaches for quantum simulations of the Fermi-Hubbard model. The first approach, based on the Hamiltonian \eqref{hamilt_hub_spin} (further referred to as ``direct method''), while the second is based on the Hamiltonian \eqref{hamilt_lgt_spin} (further referred to as ``LGT method''). It is useful to compare the corresponding computational resources required for the implementation of each of these strategies in terms of the number of qubits and CNOT gates. In order to carry out simulations for a system of $N$ sites, one needs $2N$ qubits for the implementation of the direct method and $3N$ qubits for the LGT method, where the presence of auxiliary bond spins accounts for the increase in the required number of qubits.

\begin{table}
\begin{tabular}{ |c|c|c|c| } 
 \hline
 The term & Number of terms & Cost per term & Total cost \\ 
 \hline
 Hopping & 2N & 2 & 4N  \\ 
 Interaction & N & 2 & 2N \\ 
 \hline
 Total & & & 6N\\
 \hline
\end{tabular}
\caption{Implementation cost in terms of 2-qubit gates for the direct method.}\label{cnot_table_straight}
\end{table}

\begin{table}
\begin{tabular}{ |c|c|c|c| } 
 \hline
 The term & Number of terms & Cost per term & Total cost \\ 
 \hline
 Hopping & 2N & 6 & 12N  \\ 
 Interaction & N & 2 & 2N \\ 
 \hline
 Total & & & 14N\\
 \hline
\end{tabular}
\caption{Implementation cost in terms of 2-qubit gates for the LGT-method.}\label{cnot_table_lgt}
\end{table}

In the tables \ref{cnot_table_straight} and \ref{cnot_table_lgt} we present the error estimates for the implementation of both models in terms of number of CNOT gates. One can see that the introduction of auxiliary spins in the LGT case results in a more than two-fold increase in the number of CNOTs for the corresponding circuit, due to the cost of introducing three-qubit gates. From this comparison one would naively expect that the LGT approach would perform much worse (in terms of errors) than the direct method. However, we show below that the LGT approach provides additional ways to mitigate errors on NISQ devices, which may result in its performance on par or even better than that of the direct method.

\section{Error mitigation}
The problem of dealing with errors is one of the central challenges in the task of carrying-out quantum simulations on NISQ architectures. Below we outline several error mitigation strategies which we used for the simulation of the FHM model. We benchmark their performance on a noisy simulator.

First, one can notice that both of the Hamiltonians \eqref{hamilt_hub} and \eqref{hamilt_LGT} have a number of conserved quantities. Therefore the Hilbert space in both cases can be separated into several subspaces which do not couple under the time evolution. These subspaces can be labeled by the eigenvalues of these conserved quantities. Therefore, if the initial state lies in a certain subspace, all the states obtained by the application of an ideal (noisless) time-evolution operator will remain in this \textit{physical} subspace. Of course, a general error would violate these conservation laws, which suggest the following error-mitigation strategy. Namely, we disregard any measurements outside the physical subspace~\cite{Smith2019}.

Evidently, both approaches (the direct one and the LGT) have {\it global} conservation laws as both Hamiltonians conserve the number of spin-up and spin-down fermions. In terms of Jordan-Wigner transformation this corresponds to the conservation of the net magnetization $\hat{S}^{z}_{\sigma} = \sum_j \hat{S}^z_{j\sigma}$ for  both subsystems of spins, encoding spin-up and spin-down fermions correspondingly. 

\begin{figure}[h!]
\includegraphics[width=8.0cm]{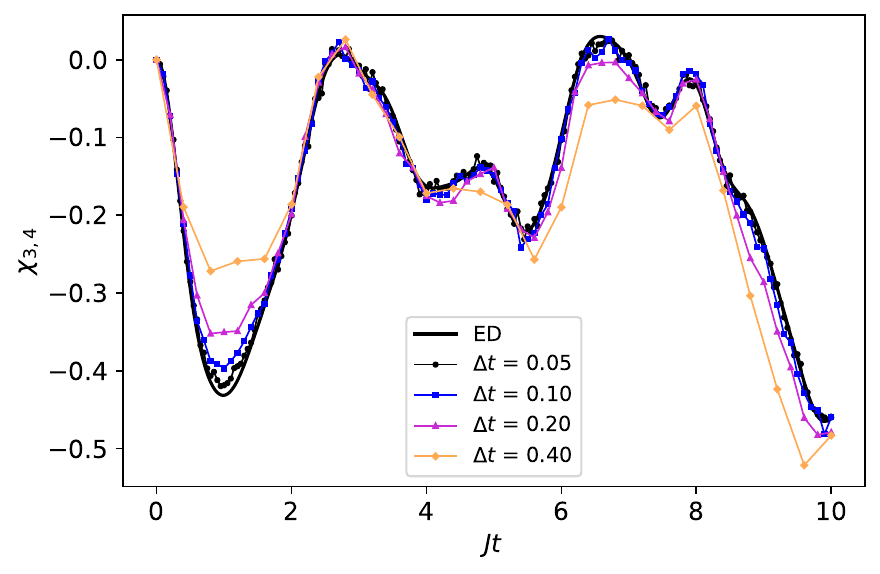}
\caption{The magnetization correlator $\chi_{34} (t)$ computed via exact diagonalization (black line), and using the LGT quantum circuit via trotterized time evolution (dots)  implemented on a Qiskit's Aer Simulator (without noise), shown for different values of a Trotter step $\Delta t$. The results are presented for a FHM on a lattice with $N = 6$ sites, and $J = 1.0$, $U = 2.0$.} \label{trotter}
\end{figure}

Our key observation is that the LGT method provides an extensive number of {\it local } conservation laws which can be used in the error mitigation. The local $\mathrm{Z}_2$ gauge invariance of the Hamiltonian \eqref{hamilt_lgt_spin} results in an {\it extensive} number $~N$ of conserved quantities, which we will denote as charges, $\hat{q}_i = (-1)^{\hat{n}_i}\hat{\tau}^{x}_{i-1,i}\hat{\tau}^{x}_{i,i+1}$, where $\hat{n}_i$ is the number of fermions on the $i$-th lattice site. These charges take on eigenvalues $\pm 1$. Therefore, due to the presence of these charges, the physical subspace for the Hamiltonian \eqref{hamilt_lgt_spin} is constrained by a larger number of conserved quantities compared with the Hamiltonian \eqref{hamilt_hub_spin}, so that, if an accidental spin-flip occurs, the probability that this measurement will be disregarded is higher.

Apart from the post-selection error mitigation, the second error mitigation strategy we employ is based on the idea that the noise in our quantum circuit depends weakly on the value of the Trotter step size $\Delta t$. We can then measure the decay of an arbitrary local operator, e.g. the number of particles or magnetization $A(t)$ on an arbitrary site with $t = n\Delta t$ for $n$ Trotter steps. We consider the Trotterized time evolution of this value, keeping the number of Trotter steps (and consequently, the depth of the corresponding quantum circuit) constant, but decreasing the Trotter step size. In the limit (without noise) $\Delta t$ going to zero the average of this operator should approach its average in the initial state. In a circuit with noise, the ratio of the measured value at time $t$ and its value in the initial state decreases for increasing number of Trotter steps (we noticed that even if $J\Delta t = 10^{-6}$, the value of $A(t)$ decays). The decay becomes independent of $t$ and we can then normalise the results of the measurements by this decay.

\section{Numerical results}

\begin{figure}[h!]
\includegraphics[width=6.0cm]{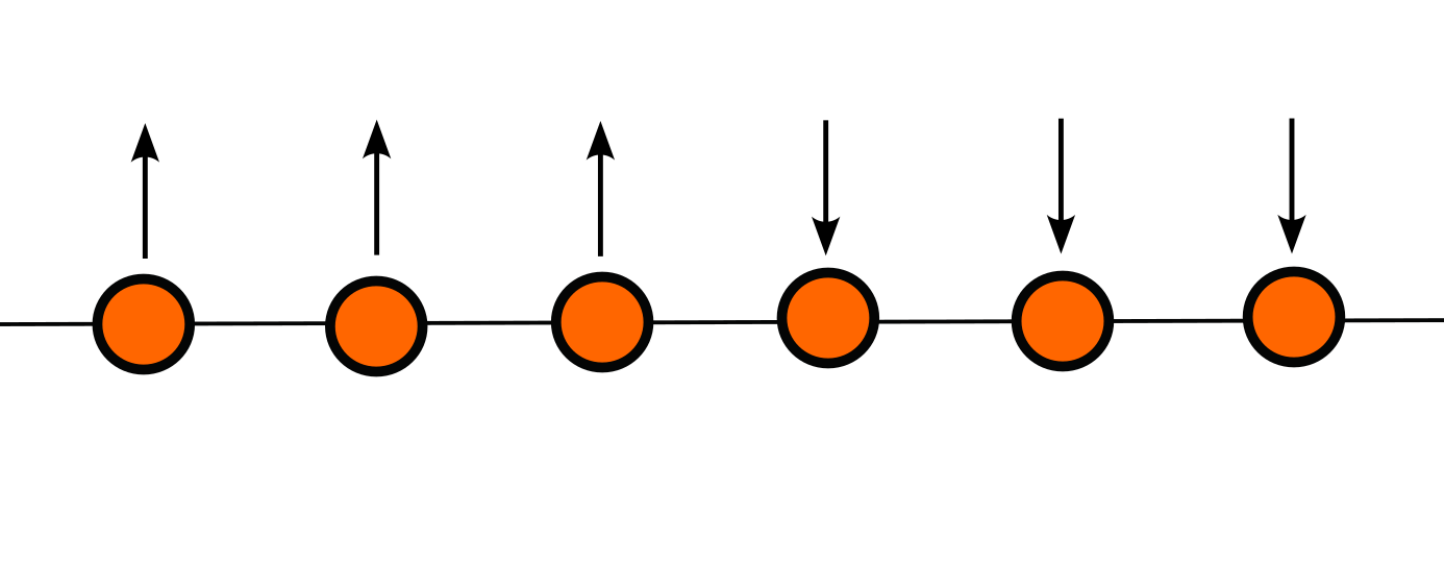}
\caption{Schematic picture of the initial state of the fermion subsystem.} \label{initial_state}
\end{figure}

In order to compare the two methods discussed above we study the time evolution of a system of spin-1/2 fermions described by the FHM model on a lattice with $N = 6$ sites. In this paper we use a classical simulator of a quantum computer with Qiskit (and compare the results of these simulations with the exact diagonalization (ED) results).
We prepare the initial state of the fermion subsystem in a domain wall configuration (see Fig.\ref{initial_state}). Namely half of the sites of the system are occupied by spin-up fermions, whereas the other half is occupied by spin-down fermions. For the bond-spin subsystem for the LGT-method, in order to make use of the advantage of having a large number of local constraints, the initial state should be chosen as an eigenstate of all charge operators $\hat{q}_i$ . In addition to that, this state should also be an eigenstate of the Wilson loop operator \eqref{gamma} Eq.(\ref{gamma}). Consequently, our initial state for the LGT-method is the tensor product $|S\rangle \otimes |\psi\rangle$ of the domain-wall state for the fermion subsystem $|\psi\rangle$, and has the following form for the bond-spin subsystem,
\be
|S\rangle = \frac{1}{\sqrt{2}}\left(|+-+-+- \rangle + |-+-+-+ \rangle\right)
\ee
Here $|+\rangle$ and $|-\rangle$ denote a bond spin polarized along the $x$-axis with the eigenvalue equal to $+1/2$ and $-1/2$ correspondingly.
\begin{figure}[h!]
\includegraphics[width=8.0cm]{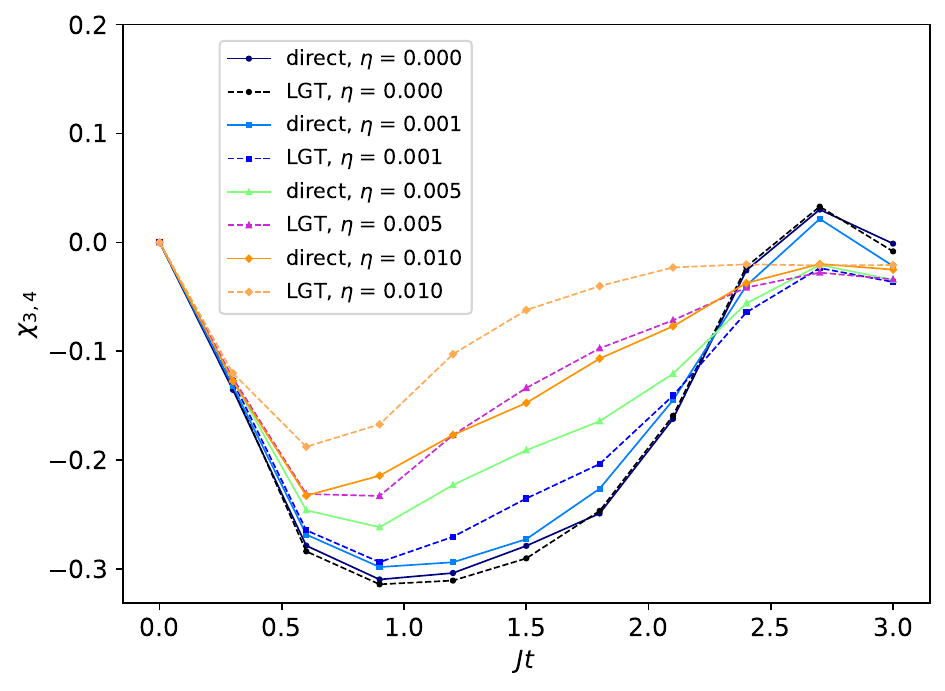}
\caption{The magnetization correlator $\chi_{3,4} (t)$ computed with the quantum circuit via trotterized time evolution which we implemented using Qiskit's QasmSimulator with different noise levels (denoted by $\eta$, see main text) without error mitigation. The results are presented for a chain of $N = 6$ sites, with $J = 1.0$ and $U = 2.0$. Trotter step $\Delta t = 0.3$.} \label{raw_var_err}
\end{figure}

\begin{figure}[h!]
\includegraphics[width=8.0cm]{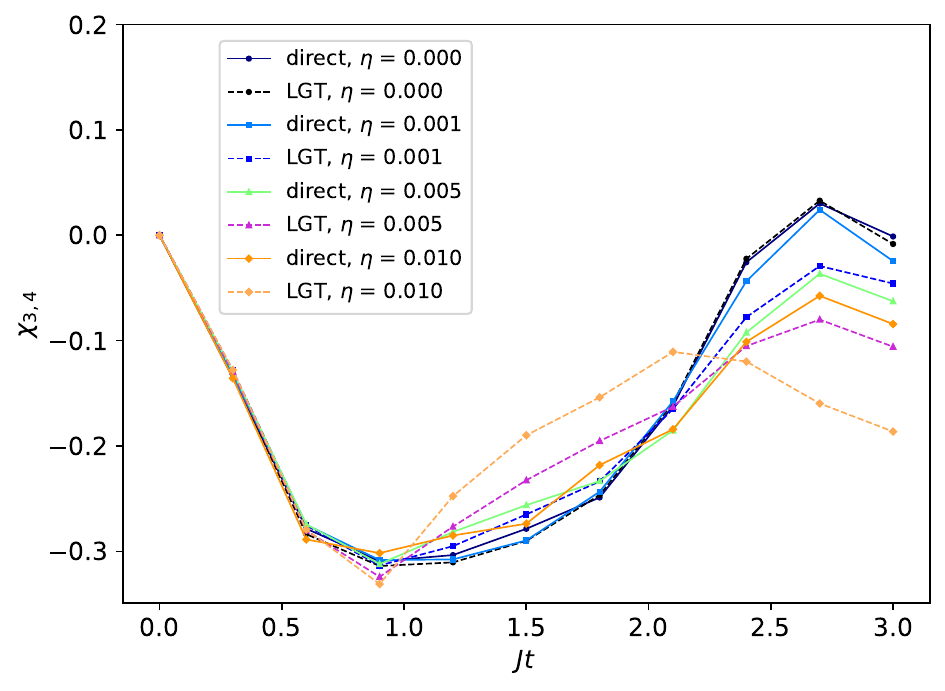}
\caption{The magnetization correlator $\chi_{3,4} (t)$ computed with the quantum circuit via trotterized time evolution which we implemented using Qiskit's QasmSimulator with different noise levels (denoted by $\eta$, see main text). These curves are shown after division by the aforementioned normalization factor $A(t)$. No post-selection was applied. The results are presented for a chain of $N = 6$ sites, with $J = 1.0$ and $U = 2.0$. Trotter step $\Delta t = 0.3$.} \label{norm_var_err}
\end{figure}

\begin{figure}[h!]
\includegraphics[width=8.0cm]{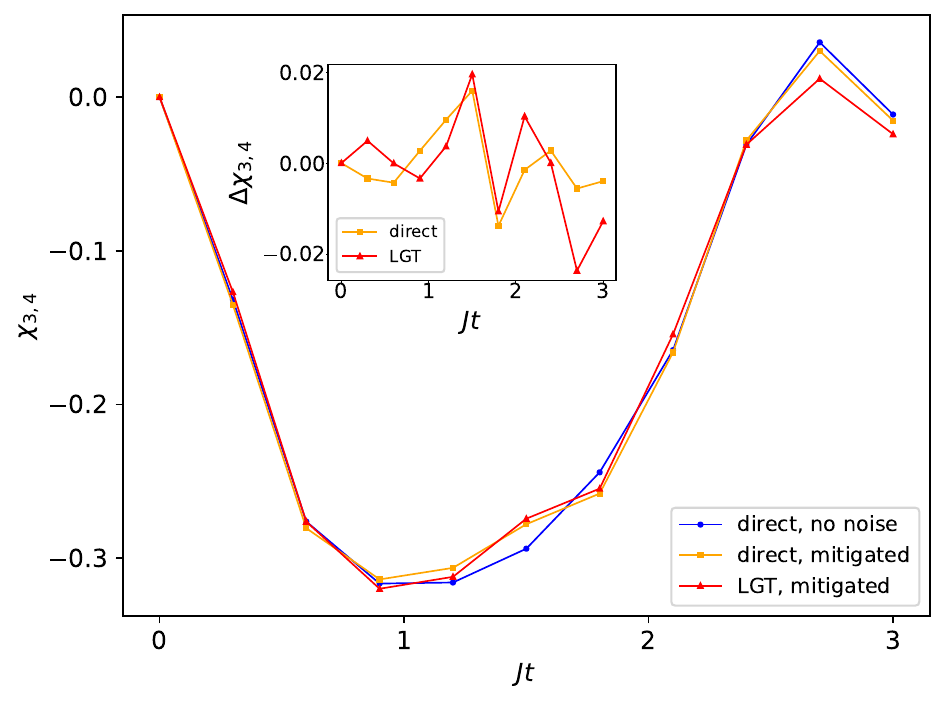}
\caption{The magnetization correlator $\chi_{3,4} (t)$ computed with the quantum circuit via trotterized time evolution which we implemented using Qiskit's QasmSimulator. The results are presented for a chain of $N = 6$ sites, with $J = 1.0$ and $U = 2.0$. Trotter step $\Delta t = 0.3$. The blue curve shows the results of an ideal machine (trotterized quantum circuit implementation with the same time-step but without noise); the red and yellow curves represent the results of noisy simulations with $\eta = 0.001$ for the LGT-method, and the direct method correspondingly. These curves are shown after our error mitigation. The inset shows the corresponding deviations from the results of an ideal machine.} \label{LGT_vs_straight_corr}
\end{figure}

\begin{figure}[h!]
\includegraphics[width=8.0cm]{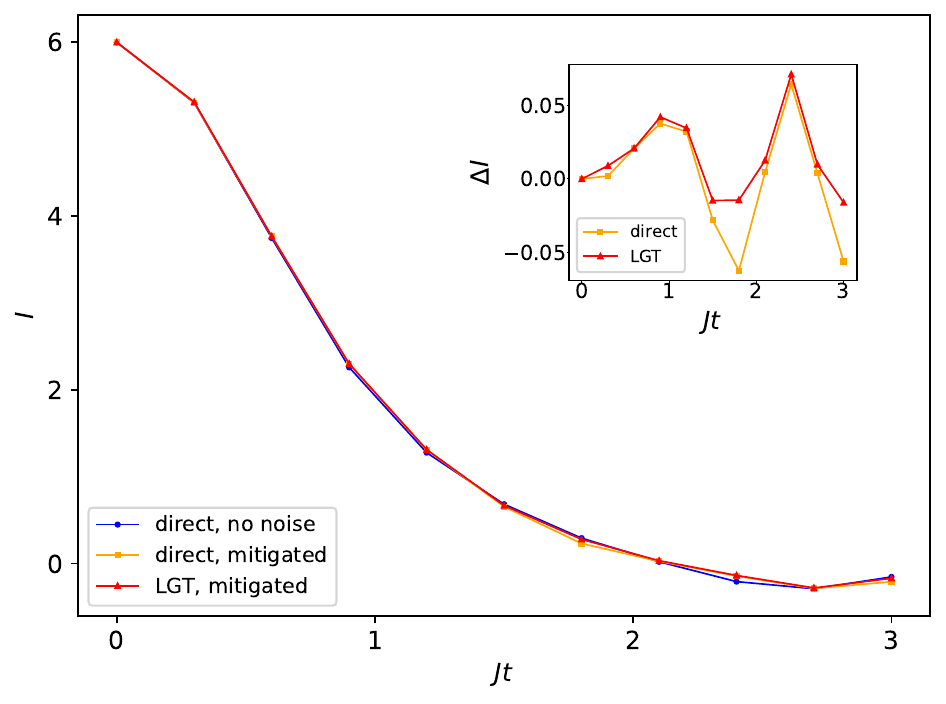}
\caption{The imbalance $I(t)$ computed with the quantum circuit via trotterized time evolution which we implemented using Qiskit's QasmSimulator. The results are presented for a chain of $N = 6$ sites, with $J = 1.0$ and $U = 2.0$. Trotter step $\Delta t = 0.3$. The blue curve shows the results of an ideal machine (trotterized quantum circuit implementation with the same time-step but without noise); the red and yellow curves represent the results of noisy simulations with $\eta = 0.001$ for the LGT-method, and the direct method correspondingly. These curves are shown after our error mitigation. The inset shows the corresponding deviations from the results of an ideal machine.} \label{LGT_vs_straight_imb}
\end{figure}

First, we show results for a noise-free ideal quantum simulator and compare the continuous time evolution calculated via ED with our LGT quantum circuit implementation for different Trotter time-steps $\Delta t$ in Fig.\ref{trotter}. Second, we study the effect of the noise on the Qiskit quantum simulator, and show how our post-selection protocol based on global and local conserved quantities can efficiently mitigate errors. We benchmark the fidelity of the numerical results by measuring a connected two-point magnetization correlator on opposite sides of the domain wall,
\be
\chi_{3,4} = \langle \hat{S}_3 \hat{S}_4 \rangle - \langle \hat{S}_3 \rangle\langle \hat{S}_4 \rangle,
\ee
where $\hat{S}_i = \hat{n}_{i,\uparrow} - \hat{n}_{i,\downarrow}$. Strong oscillations of this quantity allow one to visualize the difference in the fidelity of the results obtained via direct and LGT quantum simulation methods. In order to keep the statistical error small and roughly the same for all data points we take as many measurements as needed to obtain 10000 measurements per data point after the post-selection. In this case, the statistical error for these local correlators is around $\sim 0.01$, which is too small to be shown in the figures. It is important to note, that as the number of discarded measurements is much higher for the LGT-method, the total number of measurements required is several times higher than for the direct method.

We present the results obtained using the LGT and the direct methods for various noise levels (see figure caption) in Fig.\ref{raw_var_err}. To illustrate the influence of the noise, we use a simple noise model with depolarizing error on single and two-qubit gates. Specifically, we employ the noise model class for Qiskit Aer 0.12.2. The standard depolarizing channel is defined as follows: 
\be
E(\rho) = (1 - \lambda)\rho + \lambda\Tr\rho\frac{I}{2^n},
\ee
where $\rho$ is a density matrix, $\lambda$ is a depolarizing error parameter and $n$ is the number of qubits the channel acts on (the trace is meant as the partial trace on the number of qubits, similar the identity). Therefore, the noise level can be characterized by two depolarizing error parameters, $\eta$ for two-qubit gates and $\kappa$ for single-qubit gates. Data obtained from actual IBMQ hardware, which can be found on the IBMQ website, shows that the noise on two-qubit gates is normally an order of magnitude stronger than the noise on single-qubit gates. Therefore, we put $\kappa = 0.1\eta$  for concreteness and can tune a single noise parameter. To benchmark the accuracy of the simulations, we compare the obtained data with the results of a noise-free simulation carried out via ED. The unmitigated simulations with noise show a decay of the amplitude with increasing simulation time as errors accumulate with time. Note, as expected from our CNOT count of the two different circuits, this decay is significantly stronger for the LGT method.

Next, we show in Fig.\ref{norm_var_err} the performance of the error mitigation strategy based on the normalisation of the results by the decay factor $A(t)$, which clearly shows the improvement of the results. Of course, this error mitigation strategy does not and cannot completely cure the errors, see e.g. the results for a relatively strong noise ($\eta = 0.01$). However, the comparison with unmitigated results clearly demonstrates the efficiency of this strategy. Finally, the combination of both error mitigation strategies -- based on post-selection and the normalization -- can result in almost complete restoration of the noise-free results (see Fig.\ref{LGT_vs_straight_corr} and Fig.\ref{LGT_vs_straight_imb}), at least at short times. Remarkably, for our choice of the noise levels, the LGT-method is on par with the direct one despite having a much deeper circuit.

\section{Discussion and conclusion}

\begin{figure}[h!]
\includegraphics[width=8.0cm]{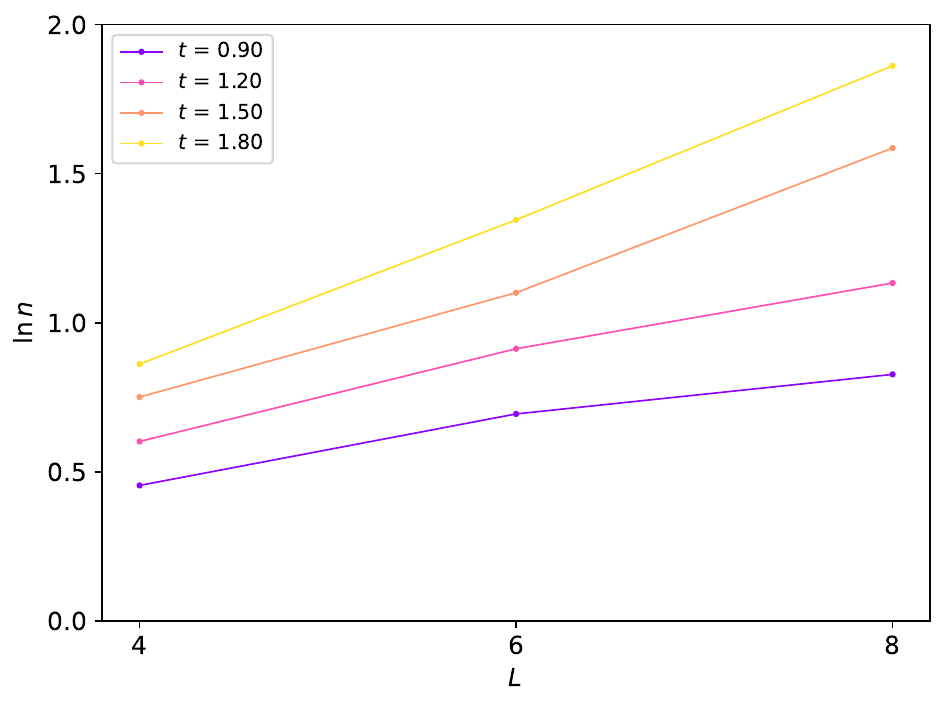}
\caption{The logarithm of the ratio between the numbers of circuit runs required for efficient post-selection in the LGT and direct simulation, respectively, as a function of system size at various points in time. Time evolution was implemented via Qiskit's QasmSimulator with the noise level $\eta = 0.005$.The results are presented for chains of $L = 4$, $6$, and $8$ sites, with $J = 1.0$ and $U = 2.0$. Trotter step $\Delta t = 0.3$.} \label{ineff_vs_L}
\end{figure}

We have presented a LGT approach for the quantum simulation of the 1D Fermi-Hubbard model on a digital quantum computer. We presented the exact mapping of the Hubbard Hamiltonian to a $\mathrm{Z}_2$ LGT, and the corresponding quantum-circuit representation for the trotterized time evolution operator. We then introduced a novel error-mitigation strategy based on the conserved quantities inherent to the LGT, which allowed us to use an efficient post-selection of measurement results obtained in the presence of noise. We compared both the LGT and the direct simulation of the model in terms of spins and find that despite a much larger number of CNOT gates required for the LGT simulation, this method may be on par with the direct implementation. It is worth noting that for increasing system sizes the number of circuit runs needed for an efficient post-selection, e.g. keeping the number of shots within the physical subspace constant, scales exponentially (see Fig. \ref{ineff_vs_L}). However, as long as runs can be parallelized and the system sizes remain small a LGT type implementation can be useful and may also profit from hardware developments for multi-qubit interactions. Moreover, to avoid the exponential scaling with system size it would be interesting to explore sampling over a fixed number of charge sectors or average charge conservation for post-selection.  Hence, our benchmark results call for further investigation of LGT-based representations of models for strongly-correlated systems for NISQ quantum computing architectures.

In our numerical simulations we implemented only incoherent noise, based on the available Qiskit simulator. A possible strategy to mitigate coherent errors could be based on the recently developed methods of linear gauge protectors and local pseudogenerators~\cite{HalimehLPG,HalimehEnhancingDFL,homeier2022quantum}. Moreover, these methods have proven to be effective in the case of incoherent $1/f$ noise~\cite{kumar2022suppression}. We expect that such a stabilization of conserved quantities of the LGT description could further increase the resilience to errors. 

The accuracy of quantum simulation results depends on two major factors, i.e. the circuit depth, and the efficiency of the error mitigation strategy. We argue that the deeper circuit doesn't always lead to less accurate results if there is an efficient and controlled way for post-selecting the data. In other words, an appropriate error mitigation strategy might be able to compensate for the lower quality of raw data. In our case, the presence of an extensive number of local conserved quantities within a LGT description of the Hubbard model makes it possible to carry out a much more stringent post-selection and to obtain better results.

\acknowledgments
We would like to acknowledge discussions with J. Halimeh, R. Moessner and A. Smith, and  related collaborations on $Z_2$ LGTs. D.K. acknowledges support from Labex MME-DII grant ANR11-LBX-0023, and funding under The Paris Seine Initiative EMERGENCE programme 2019. J.K. acknowledges support from the Imperial-TUM flagship partnership, as well as the Munich Quantum Valley, which is supported by the Bavarian state government with funds from the Hightech Agenda Bayern Plus. DK and JK Acknowledge support provided by the French-German ANR-DFG (German Research Foundation) grant 505662248.

\bibliography{biblio_hubbard}

\end{document}